\documentclass{article}

\usepackage{arxiv}

\usepackage[utf8]{inputenc} %
\usepackage[T1]{fontenc}    %
\usepackage{url}            %
\usepackage{booktabs}       %
\usepackage{amsfonts}       %
\usepackage{nicefrac}       %
\usepackage{microtype}      %
\usepackage{lipsum}
\usepackage{graphicx}
\usepackage{caption}
\usepackage[backend=biber]{biblatex}

\title{CBIR using features derived by Deep Learning}

\author{
  Subhadip Maji\thanks{GitHub Repo: \emph{https://github.com/pidahbus}} \\
  M.Tech QROR-II \\
  Indian Statistical Institute, Kolkata \\
  Kolkata, 700108 \\
  \texttt{qr1705@isical.ac.in} \\
   \And
 Smarajit Bose \\
  Interdisciplinary Statistical Research Unit \\
  Indian Statistical Institute, Kolkata \\
  Kolkata, 700108 \\
  \texttt{smarajit@isical.ac.in} \\
}
\DeclareUnicodeCharacter{2212}{-}
\begin{document}
\maketitle

\begin{abstract}
In a Content Based Image Retrieval (CBIR) System, the task is to retrieve similar images from a large database given a query image. The usual procedure is to extract some useful features from the query image, and retrieve images which have similar set of features. For this purpose, a suitable similarity measure is chosen, and images with high similarity scores are retrieved. Naturally the choice of these features play a very important role in the success of this system, and high level features are required to reduce the “semantic gap”. 

In this paper, we propose to use features derived from pre-trained network models from a deep-learning convolution network trained for a large image classification problem. This approach appears to produce vastly superior results for a variety of databases, and it outperforms many contemporary CBIR systems. We analyse the retrieval time of the method, and also propose a pre-clustering of the database based on the above-mentioned features which yields comparable results in a much shorter time in most of the cases.

\end{abstract}


\section{Introduction}

Given a query image, often similar images may need to be retrieved from a large database. This is called Content Based Image Retrieval. The standard procedure is to find similar images based on some features extracted from the images. Ideally these features should describe the content information of the images. That is why high-level features are needed, and the low level features like pixel values etc are not very useful. 

IBM developed the first commercial version of CBIR system naming QBIC (Query By Image Content)\cite{niblack} in 1995. It allows user to query by user-constructed sketches, example images and drawings. This system uses a combination of texture, shape and colour. Colour co-occurrence matrix (CCM) is used to extract low level features from images, which has been widely used in many works\cite{shim, huang, ojala} in the area of CBIR. Bose et al. \cite{bose} used some visual descriptors to extract features from MPEG-7 standard \cite{kosch, manjunath}  along with CCM features which achieved some improvement. 

Lohite et al. \cite{lohite} worked with the widely used color, texture and edge features of the images, and optimized the result using SVM (Support Vector Machine) classifier. Mehmood et al.\cite{Mehmood} presented a CBIR method named WATH (weighted average of triangular histograms) of visual words. This method adds image spatial elements to inverted index of BoVW (bag-of-visual-words) model, corrects overfitting problem on larger size of dictionary and tries to bride the semantic gap between low-level and high-level features. 

Rashno et al.\cite{rashno} proposed a new scheme which suggests to transform the input RGB image to three subsets in neutrosophic (NS) domain. For each of the segment, statistic component, histogram, colour features including dominant colour descriptor (DCD) and wavelet features are extracted. These features are then used to retrieve images. 

Kumar et al.\cite{SatyaKumar2018} introduced a new feature descriptor called local mean differential excitation pattern (LMDeP) which can produce robust features. Sarwar et al.\cite{sarwar} recommended a method based on bag-of-words (BoW) model, which integrates visual words with local intensity order pattern (LIOP) feature and local binary pattern variance (LBPV) feature to reduce the semantic gap issue and enhance CBIR performance. Rana et al.\cite{rana} proposed image retrieval by combining colour and shape features with nonparametric ranklet transformed texture features. Yusuf et al.\cite{yousuf} gained improvement in CBIR performance on the basis of visual words fusion of scale invariant feature transform (SIFT) and local intensity order pattern (LIOP) descriptors. Sharif et al.\cite{sharif} came up with another feature descriptor called binary robust invariant scalable keypoints (BRISK) along with SIFT. Ashraf et al.\cite{ashraf} developed a method which retrieves images using YCbCr colour scheme with canny edge histogram and discrete wavelet transform. In Obulesu et al.’s \cite{obulesu} two extended versions of motif co-occurrence matrices (MCM) are calculated and combined to improve the CBIR performance.

After the introduction and evolution of Deep Learning Neural Network, the performance of CBIR has got a boost, because by the help of deep models we can finally extract higher-level features along with the low-level features from the image to reduce the semantic gap mentioned above. Khokhar et al.\cite{khokhar} described how Back-propagation Feedforward Neural Network (BFNN) can be used for classification in CBIR  after exploiting some features of images e.g. geometric, colour and texture. Ashraf et al.\cite{ashraf2} presented a bandlet transform based image representation technique which returns information about major objects present in the image reliably. Finally to retrieve images Artifical Neural Network has been used. Xu et al.\cite{jian} proposed part-based weighting aggregation (PWA) for CBIR. This PWA utilizes discriminative filters of deep convolutional layers as part detectors. Several other recent CBIR techniques can be found in the review paper\cite{zhou}.

In this paper, features derived from a pre-trained network model from a deep learning convolution neural network trained for a large image classification have been used for retrieval of similar images. The resulting algorithms appears to achieve remarkable success in terms of retrieval accuracy, and appears to outperform many contemporary CBIR methods. The algorithm is quite fast, however, to reduce retrieval time, a concept of pre-clustering the database has also been introduced which seems to work faster without sacrificing retrieval performance.

The paper is organized is as follows: Section 2 describes the motivation behind this approach, presents a review of the key ingredients and explores the characteristics of the derived features from a pre-trained network model. In Section 3, we present the details regarding the pre-trained models, similarity measures
and evaluation of performance that have been used in this work. Section 4 contains the results of extensive experiments while Section 5 discusses the time complexity. In Section 6, we introduce a concept of pre-clustering of the database and in Section 7, we present the concluding remarks and some future directions.

\section{Proposed Method using Features derived by Deep Learning}
\label{sec:headings}

In a Content-Based Image Retrieval system, images are represented by a set of low level or/and high-level features. This is called feature encoding where an image from RGB or HSV space is encoded to a n-dimensional feature vector. In this paper we propose to derive feature vectors of an image with the help of some pre-trained deep learning models. For this purpose we first present a brief description of the key concepts such as neural network, pre-trained models etc. 

\subsection{Neural Network}
Neural Networks are set up as collections of neurons that are connected in a non-cyclic graph. These models are often depicted into separate layers of neurons. Generally, the most common type is the fully-connected Neural Network layer in which neurons between two adjacent layers are fully pairwise connected, but there is no connection between neurons within a single layer. Below are two examples of fully connected Neural Network\cite{stanford}.

\begin{figure}[htp]
    \centering
    \includegraphics[scale=0.8]{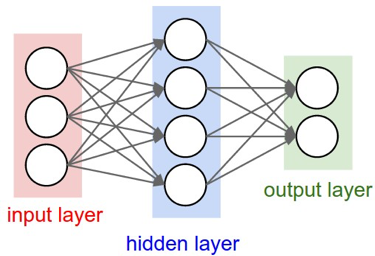}
    \includegraphics[scale=0.8]{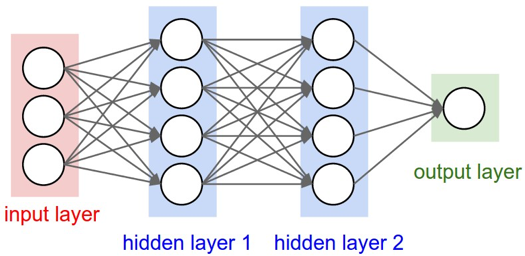}
    \caption{Left: A two-layer Neural Network (one hidden layer of four neurons and one output layer with two neurons) with three inputs. Right: A three-layer neural network with two hidden layers of four neurons each, one output layer with a single neuron and three inputs \cite{stanford}}
    \label{fig:densenn}
\end{figure}

\subsection{Convolutional Neural Network (CNN)}
\begin{figure}[htp]
    \centering
    \includegraphics[scale=0.8]{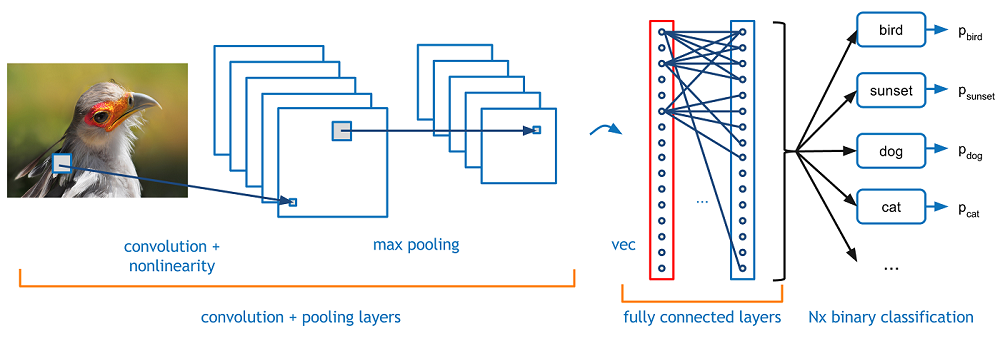}
    \caption{Every layer of a CNN converts the 3D input volume to a 3D output volume of activations. In given example, the image of the bird is the input layer, so its height and width are the dimensions of the image, and the depth would be three (Red, Green, Blue channels)}
    \label{fig:cnn}
\end{figure}

Convolutional Neural Networks take input as images and they handle the architecture in a more sensible way. The layers of a CNN have neurons which are arranged in 3 dimensions: width, height, depth. Here, the word depth refers to the third dimension of an activation volume of a layer. The neurons in a layer are connected to a small region of the preceding CNN layer, unlike to all the neurons which is a norm in a fully-connected  neural network. The visualization is shown in Figure \ref{fig:cnn}.

As mentioned above, a simple CNN is a sequence of layers, and every layer of a CNN transforms one volume of activation to another by passing through certain differentiable functions. There are mainly three types of layers in CNN architectures: Convolutional Layer, Pooling Layer and Fully-Connected Layer (shown in Figure \ref{fig:densenn}). We will stack these layers on top of each other to form a fully operational CNN architecture\cite{stanford}. Figure 2 shows the CNN model architecture of a classification problem. This network consists of convolution, pooling, fully connected layers and some activation layers (e.g. ReLU, softmax etc).

\subsection{Pre-trained Neural Network Model}
Transfer learning\cite{torrey} is a method where instead of starting the training process of a model from scratch, we use the learned weights of an already trained model to solve a different but similar problem. In this way we leverage previous learning through the learned weights and save time considerable amount of time. Usually much better results are also achieved compared to training from scratch. 

In computer vision, transfer learning is usually executed by the use of pre-trained models. A pre-trained model\cite{pre-trained} is a model that was trained on a large benchmark dataset to solve some specific problems similar to the ones we want to solve. Accordingly, because of the high computational cost of training such deep learning models, it is a common practice to import and use models from a published architecture (e.g. VGG, ResNet, Xception etc). A comprehensive review of pre-trained models’ performance on computer vision problems using data from the ImageNet\cite{deng} challenge is presented by Canziani et al.\cite{canziani}.

Being motivated by this, we have used a pre-trained Neural Network model which was trained on the  ImageNet Dataset. This dataset contains more than 14 million images which belong to more than 20,000 classes. It also provides bounding box annotations for around 1 million images, which can be used in Object Localization tasks. Multiple layers of convolutional layers, average pooling layers, max pooling layers etc. are stacked up with one another with different combinations to build up the model architectures. The final layer is then unwrapped to an $n$-dimensional vector, which is called the dense (or fully connected) layer. Several fully connected layers can be stacked on this unwrapped dense layer. Finally, a softmax activation layer of dimension: the number of class is stacked over the final dense layer to get the probabilities of each class for classification problems or a linear activation layer of single dimension can be placed to predict the regressed value for regression problems or other kind of structured activation layers are placed according to the desired problems to solve. 

\subsection{Visualizing what CNN learn}
It is often referred that deep-learning models are “black boxes”. For certain types of deep-learning models it may be true but for CNN it is not absolutely true. The representations and features, learned by CNN are highly amenable to visualization, in large part because they’re representations of visual concepts. Since 2013, different types of techniques have been developed for visualizing and interpreting these feature representations of CNN. Below are some methods to visualize the learnings of CNN\cite{Chollet}.

\subsubsection{Visualizing intermediate CNN outputs (intermediate activations):}

\begin{figure}[htp]
    \centering
    \includegraphics[scale=0.8]{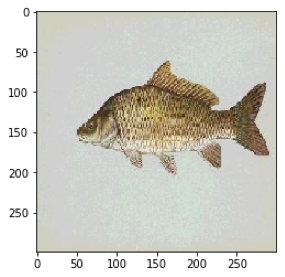}
    \caption{A sample image from DB2000 Dataset}
    \label{fig:fish1}
\end{figure}

\begin{figure}[htp]
    \centering
    \includegraphics[scale=0.8]{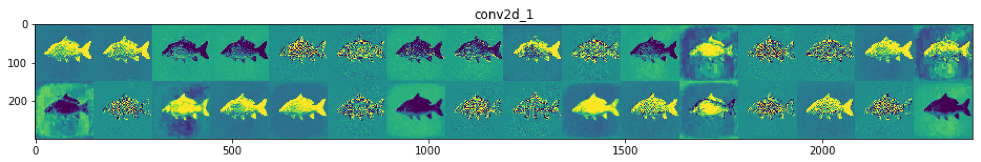}
    \includegraphics[scale=0.8]{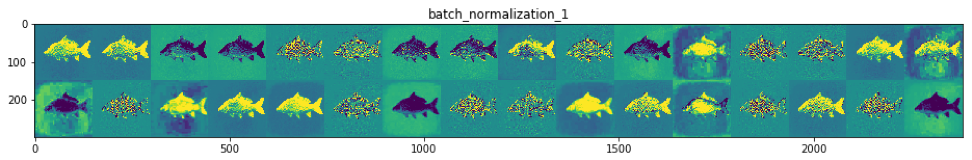}
    \includegraphics[scale=0.8]{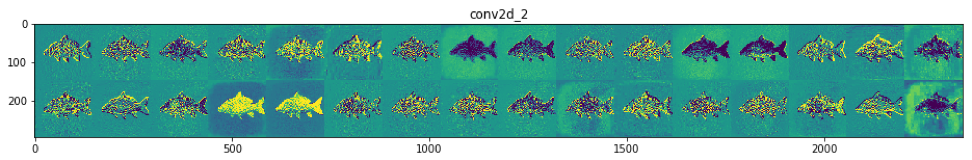}
    \caption{First few intermediate activations of the image shown in Figure \ref{fig:fish1}}
    \label{fig:activation1}
\end{figure}

\begin{figure}[htp]
    \centering
    \includegraphics[scale=0.8]{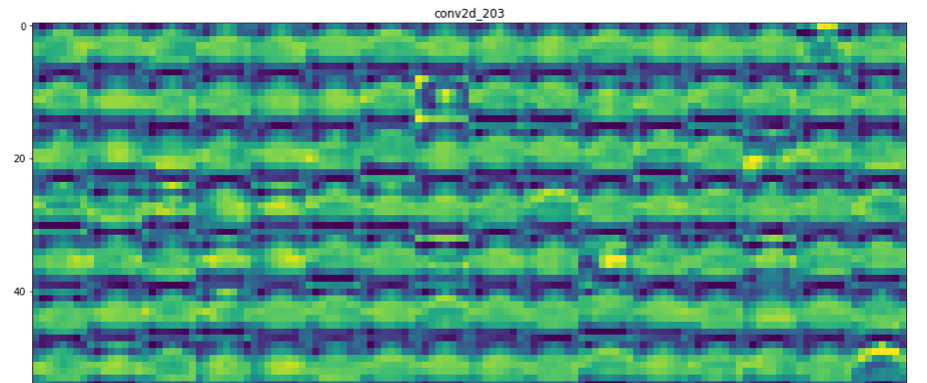}
    \includegraphics[scale=0.8]{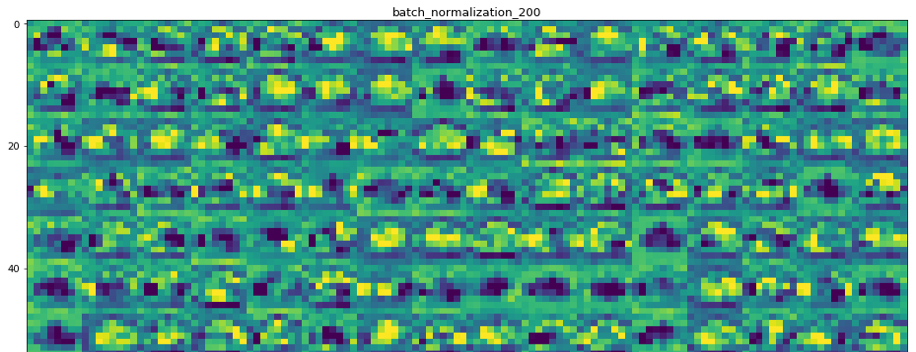}
    \caption{Last few intermediate activations of the image shown in Figure \ref{fig:fish1}}
    \label{fig:activation2}
\end{figure}

For this sample image shown in Figure \ref{fig:fish1}: from ImageDB2000 Dataset, the first few intermediate activations for the above image are shown in the Figure \ref{fig:activation1}. The last few intermediate layer activations for Figure \ref{fig:fish1} are shown in the Figure \ref{fig:activation2}.

There are a few things to note here:
\begin{enumerate}
    \item The first layers are basically the edge detectors. At this stage, the activations retain almost all of the information present in the picture fed in the network.
    \item As we go higher in the model, the activations outputs from each layer become increasingly abstract and less visually interpretable. They begin to encode higher-level features such as “fish fin” and “fish eye.” Higher feature representations carry increasingly less information about the visual contents of the image, and increasingly more information related to the class of the image.
    \item The sparsity of the activations increases as we go deep in the layers of CNN.
\end{enumerate}

\subsubsection{Visualizing intermediate convnet outputs (Adaptive Deconvolutional Networks)}

\begin{figure}[htp]
    \centering
    \includegraphics[scale=0.8]{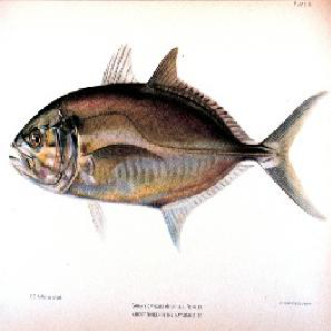}
    \caption{A sample image of Fish}
    \label{fig:fish2}
\end{figure}

\begin{figure}[htp]
    \centering
    \includegraphics[scale=0.8]{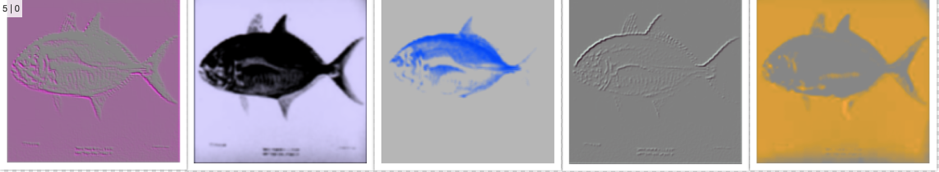}
    \caption{Some low level features of the image shown in Figure \ref{fig:fish2}}
    \label{fig:shallow}
\end{figure}

\begin{figure}[htp]
    \centering
    \includegraphics[scale=0.8]{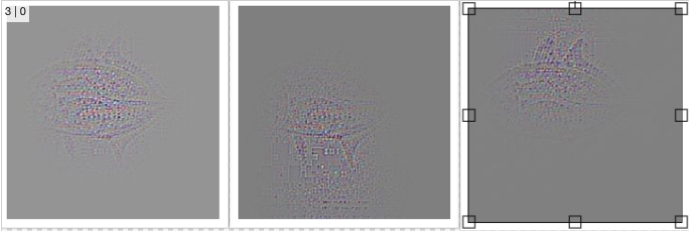}
    \caption{Some of the deep level features of the image shown in  Figure \ref{fig:fish2}}
    \label{fig:deep}
\end{figure}

This model produces an over-complete image feature representation that can be used as input to standard neural network object classifiers. This model is learned from natural images and, given a new image, requires inference to compute. It decomposes an image in a hierarchical fashion using multiple alternating layers of convolutional sparse coding (deconvolution\cite{zeiler}) and max-pooling. Each of this deconvolution layers attempts to minimize the reconstruction error of the input image under a sparsity constraint on an over-complete set of feature maps. After doing so, for this sample image shown in Figure \ref{fig:fish2}. Some of the shallow level (low level) convolution features are shown in Figure \ref{fig:shallow}. And, some of the deep level (high level) convolution features are shown in Figure \ref{fig:deep}.

Here we actually see that the model represents edge, texture type low-level features in the first layers, where in the last layers the model learns to represent some higher-level features. As an example, in Figure \ref{fig:deep} the deep layers of the model represents fish fin, fish body types concepts.

Visualizing convent filters has been clearly described in chapter 5 of Francois Chollet’s book\cite{Chollet}. It has been shown that the filters in the earlier layers encode directional edges, colors and textures. Also, as we visualize the filters in the deeper layer we find that the filters lend to learn textures found in natural images: feathers, eyes, leaves etc.

\subsection{Proposed method of Feature Extraction by Deep Learning}

From the above section it is evident that as the model architecture goes deep, it starts to learn high-level features from the low-level features. We propose to use these higher-level features for the feature representation of images in a CBIR system. So, we removed the last softmax activation layer used for calculating probabilities of each class and selected the preceding fully connected layer to be our feature vector representation for CBIR. As this vector is the deepest layer of the model, this represents the most learned high-level features. We encoded (predicted) the images of our CBIR database through our pre-trained model and got an $n$-dimensional feature vector for each of the images. The flowchart of this process is shown in Figure \ref{fig:flowchart1} for better understanding. The value of $n$ varies with the selection of deep learning network architecture. 

\section{Details of the proposed algorithm and performance evaluation}

\subsection{Pre-trained models used}

We have tried the following network architectures all pre-trained on the ImageNet dataset: 
\begin{itemize}
    \item DenseNet\cite{densenet}
    \item InceptionResNetV2\cite{inception}
    \item InceptionV3\cite{inception}
    \item MobileNetV2\cite{mobilenetv2}
    \item NasNet Large\cite{nasnet}
    \item ResNet50\cite{resnet}
    \item VGG19\cite{vgg}
    \item Xception\cite{xception}
\end{itemize}

The main advantage of this method of feature extraction is that now we are able to extract higher level features without exploiting our database. This is necessary because from the practical point of view we would be having a dump of dataset without any class information. The user will try to find similar images from the database based on the query image he/she has. Now as the dump dataset does not have any pre-defined class, training model on our dataset is not a feasible task unless we manually try to assign class to each of the images of our database dump consisting millions or billions of images which is very time consuming and prone to subjective error for classifying images. To avoid this problem, we are using a Neural Network model pre-trained on a huge separate dataset (ImageNet) to perform feature extraction independently on our CBIR datasets unlike training the model itself on the CBIR datasets\cite{khokhar, jabeen}

\subsection{Database Used}
The CBIR methods were applied on the following image databases, which vary in number as well as types of images.

\begin{itemize}
    \item \textbf{ImageDB2000:} The database contains 2000 images from 10 different categories each containing 200 images. The categories are Flowers, Fruits, Nature, Leaves, Ships, Faces, Fishes, Cars, Animals, and Aeroplanes\cite{bose}. 
    
    \item \textbf{ImageDBCaltech (Caltech101):} This database contains 9144 images from 102 categories. The number of images in each category varies from 34 to 800\cite{fei-fei}.
    
    \item \textbf{ImageDBCorel:} This dataset contains 1000 images belonging to 10 categories. Each category contains 100 images. The categories are: African People, Beach, Building, Bus, Dinosaurs, Elephant, Flower, Horse, Mountain and Food\cite{corel}.
\end{itemize}

\begin{figure}[htp]
    \centering
    \includegraphics[scale=0.7]{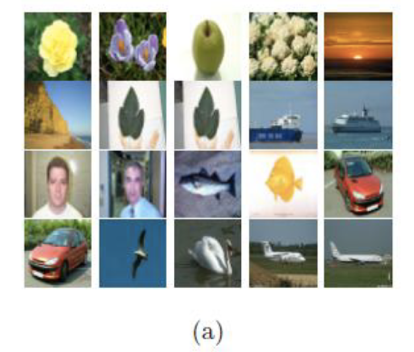}
    \includegraphics[scale=0.7]{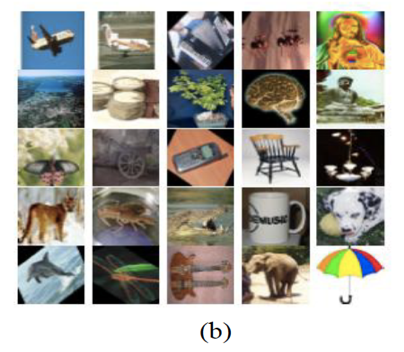}
    \caption{Sample images from Dataset (a) ImageDB2000 and (b) DBCaltech}
\end{figure}

\begin{figure}[htp]
    \centering
    \includegraphics[scale=0.6]{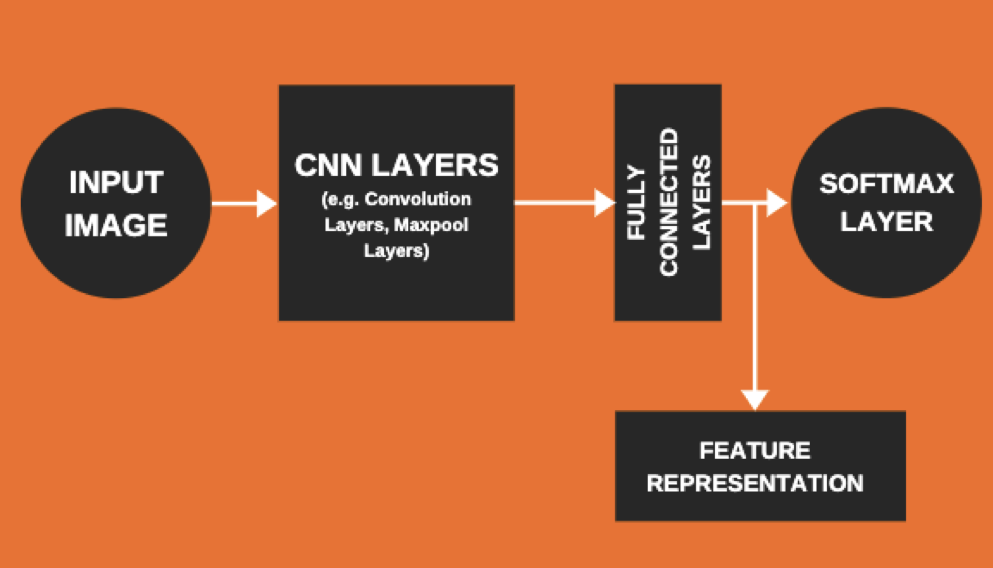}
    \caption{CBIR image feature representation with the help of pre-trained deep learning models}
    \label{fig:flowchart1}
\end{figure}

\subsection{Similarity Measure}
The similarity (or dissimilarity) between a query image (Q) given by the user and a database image (I) stored in the system is measured by some distance metric. It is assumed that this distance will accurately measure the dissimilarity (or similarity) between the images as well. A smaller calculated distance implies more similarity. The similarity between two images can be different for different user’s way of perceiving the images. Broadly speaking, there are mainly two types of similarity measures, geometric and probabilistic\cite{Peng}. In the first case, the similarity is based on the distance between the feature vectors. A most widely used one is Minkowski, of which L1−norm and L2−norm are most popular. In probabilistic type, a Gaussian classifier is often used to measure the relevance between the query image and the database image so that the pairs that had a high likelihood ratio were classified as relevant and the pairs having a low likelihood ratio is considered as irrelevant. Although past research\cite{aksoy} shows that the probabilistic methods perform significantly better than the geometric methods, they are computationally expensive. Throughout this paper we have used L1 or L2 norm as dissimilarity measure.

Manhattan Distance (L1 norm) between the extracted features of query image (Q) and database image (I) is formulated by,

\begin{equation}
    D(I,Q) = \sum_{i=1}^{n} |x_{i,I} - x_{i,Q}|
\end{equation}

And, Euclidean Distance (L2 norm) between Q and I is given by,

\begin{equation}
    D(I,Q) = \sum_{i=1}^{n} (x_{i,I} - x_{i,Q})^2
\end{equation}

Where, n is the feature dimension of the images. $x_(i,I)$ and $x_(i,Q)$ are the i-th feature value of database image and query image respectively.

\subsection{Evaluation of Performance}
The most commonly used measures for evaluating the performance of a CBIR system is Precision, which is defined as follows:

\begin{equation}
    \texttt{Precision} = \frac{\texttt{Number of relevant images retrieved}}{\texttt{Number of retrieved images}}
\end{equation}

Generally, the number of images retrieved by any CBIR method is a pre-specified positive integer. This is called the scope of the system. Precision value is calculated for each image in the database, and these values are averaged over all images in the database. Usually, the greater the scope, the larger is the number of relevant images retrieved, typically leading to decreasing values of precision.

\section{Results}

In this section we present all the results including selection of the best deep learning network architecture, retrieval results of our CBIR system with respect to the sample query images taken from our datasets, category wise image retrieval precision for our datasets and comparison of precision between the proposed method and some other recent ones on CBIR. 

\subsection{Selection of best Deep Learning Network architecture}

\begin{figure}[htp]
    \centering
    \includegraphics[scale=0.7]{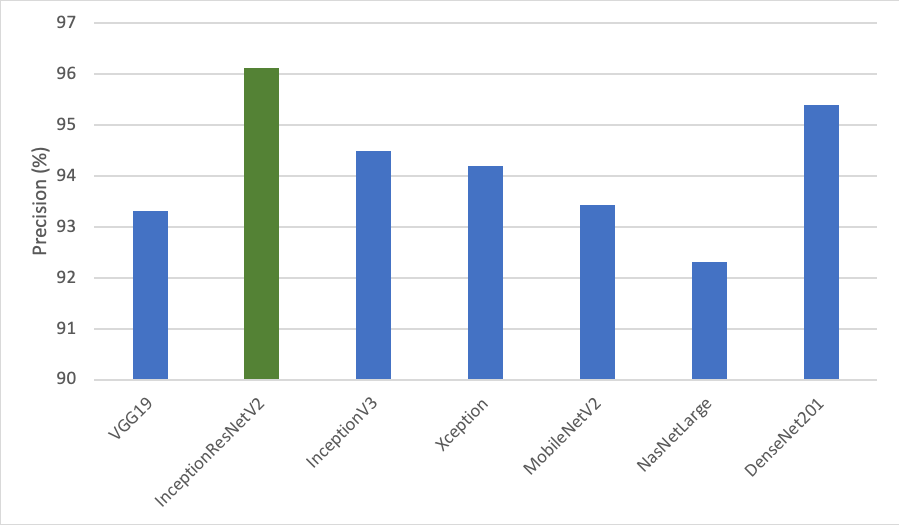}
    \caption{Comparison of Deep Learning Architecture on Corel Dataset}
    \label{fig:comp_arch}
\end{figure}

 We did a comparative study to select the best deep learning architecture as described in section 2.5 by calculating the average precision for each one of them for a scope value of 20 on DBCorel dataset. Euclidean Distance (L2 norm) is used as the dissimilarity metric. From the results given in Figure \ref{fig:comp_arch} it is clearly seen that InceptionResNetV2 is the winner with 96.115\% average precision value. Therefore we decided to use the InceptionResNetV2 network architecture for all the subsequent experiments.

\subsection{Image Retrieval of Sample Query Images}

\begin{figure}[htp]
    \centering
    \includegraphics[]{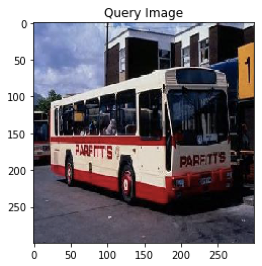}
    \caption{A sample query image from DBCorel}
    \label{fig:sample1_corel}
\end{figure}

Using the InceptionResNetV2 architecture on the Corel Dataset for the scope of 20, for the example query image shown in Figure \ref{fig:sample1_corel}.

\begin{figure}[htp]
    \centering
    \includegraphics[scale=0.7]{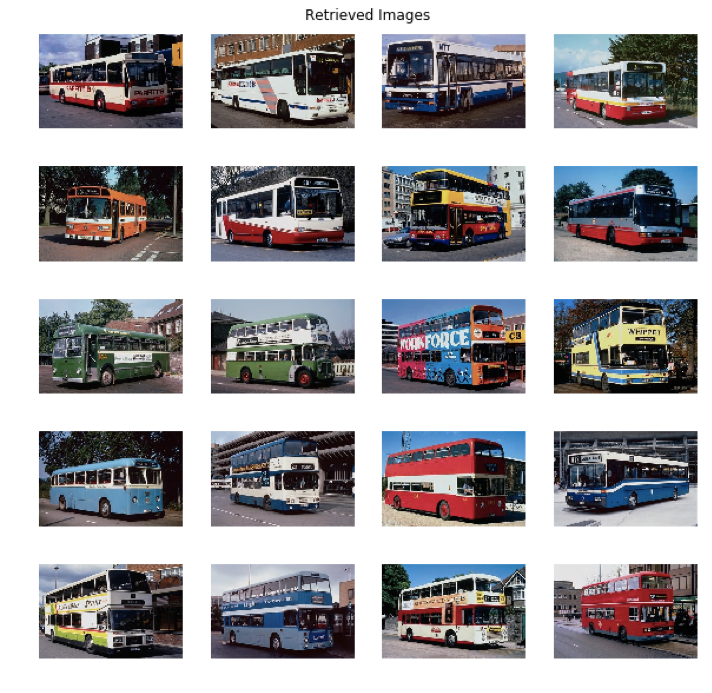}
    \caption{Retrieved Results for the query image shown in Figure \ref{fig:sample1_corel}}
    \label{fig:retrieved1_corel}
\end{figure}

We retrieve 20 results shown in Figure \ref{fig:retrieved1_corel}. From Figure \ref{fig:retrieved1_corel}, we find that the query images belong to the “bus” category and all 20 results are relevant to the query image. So, the precision for this query image is 1.

\begin{figure}[htp]
    \centering
    \includegraphics[]{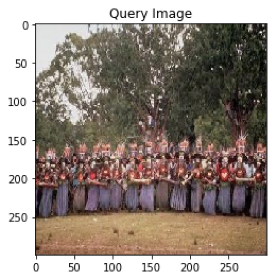}
    \caption{A sample query image from DBCorel}
    \label{fig:sample2_corel}
\end{figure}

For another query image from the corel dataset shown in Figure \ref{fig:sample2_corel}, the retrieved results are shown in Figure . Here we see that the query image belongs to the “African People” category and out of 20 retrieved results 13 results are relevant to the query image. So, for this specific image precision value is 13/20 = 0.65. 

\subsection{Category wise Precision Calculation}

In this subsection we produce the category wise average precision on DBCorel and DB2000 for a scope of 20 using Euclidean Distance as dissimilarity metric. Fig \ref{fig:categorywise} and Table \ref{tab:2000_corel_cat} illustrate the results. From the result we see that in DBCorel, for Bus, Dinosaurs and Elephant category the pre-trained InceptionResNetV2 model retrieves all the images with 100\% precision but performs comparatively poorly for the African People category resulting in 79.35\% precision. The overall average precision for this dataset is 96.115\%. For DB2000 the best retrieved category is Airplane (precision: 99.975\%) and worst category is Leaf (precision: 91.125\%), overall average precision being 97\%. 

\begin{figure}[htp]
    \centering
    \includegraphics[scale=0.7]{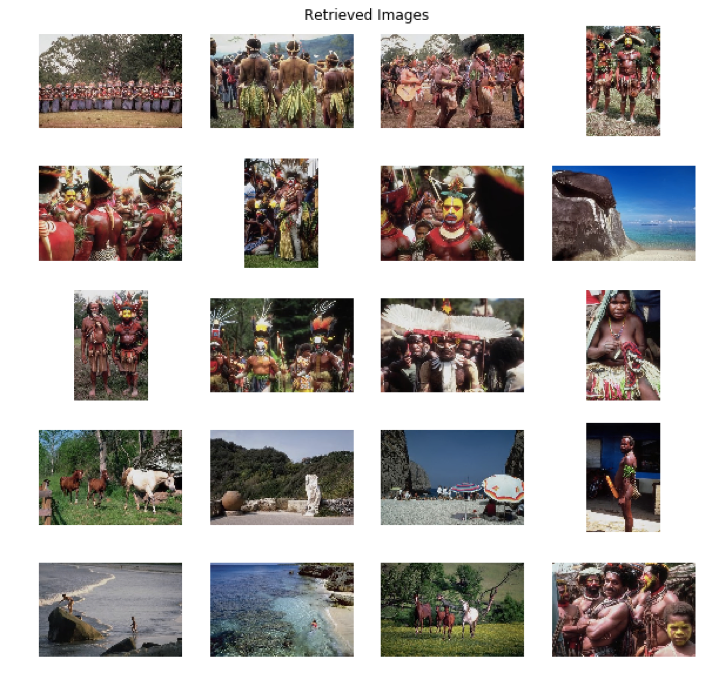}
    \caption{Retrieved Results for the query image shown in Figure \ref{fig:sample2_corel}}
    \label{fig:retrieved2_corel}
\end{figure}

\begin{figure}[htp]
    \centering
    \includegraphics[scale=0.8]{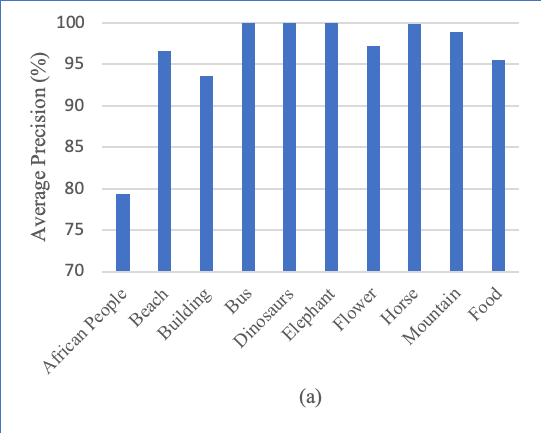}
    \includegraphics[scale=0.8]{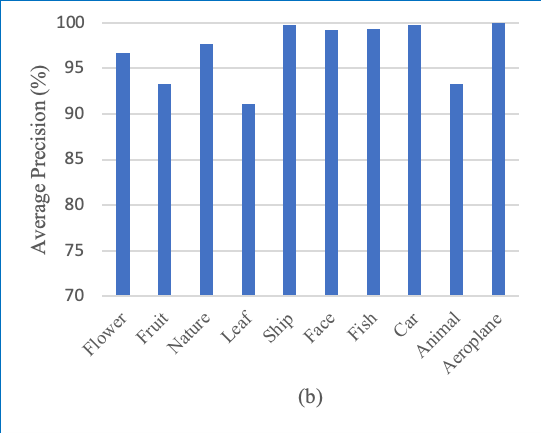}
    \caption{Category wise average precision for scope of 20 on (a) DBCorel (b) DB2000}
    \label{fig:categorywise}
\end{figure}

\begin{table}[ht]
    \begin{minipage}[b]{0.45\linewidth}\centering
        \begin{tabular}{|c|c|c|}
            \hline
            \textbf{Categories} & \textbf{Average Precision(\%)}\\
            \hline
            African People & 79.35\\
            \hline
            Beach & 96.6\\
            \hline
            Building & 93.55\\
            \hline
            Bus & 100\\
            \hline
            Dinosaurs & 100\\
            \hline
            Elephant & 100\\
            \hline
            Flower & 97.25\\
            \hline
            Horse & 99.9\\
            \hline
            Mountain & 98.95\\
            \hline
            Food & 95.55\\
            \hline
        \end{tabular}
    \caption*{(a)}
    \end{minipage}
    \hspace{0.5cm}
    \begin{minipage}[b]{0.45\linewidth}
    \centering
        \begin{tabular}{|c|c|c|}
            \hline
            \textbf{Categories} & \textbf{Average Precision(\%)}\\
            \hline
            Flower & 96.65\\
            \hline
            Fruit & 93.25\\
            \hline
            Nature & 97.675\\
            \hline
            Leaf & 91.125\\
            \hline
            Ship & 99.8\\
            \hline
            Face & 99.275\\
            \hline
            Fish & 99.3\\
            \hline
            Car & 99.725\\
            \hline
            Animal & 93.3\\
            \hline
            Aeroplane & 99.975\\
            \hline
        \end{tabular}
    \caption*{(b)}
    \end{minipage}
    \caption{Category wise average precision for scope of 20 on (a) DBCorel (b) DB2000}
    \label{tab:2000_corel_cat}
\end{table}

\subsection{Result comparison with other recently proposed algorithms}

\begin{figure}[htp]
    \centering
    \includegraphics[scale=0.7]{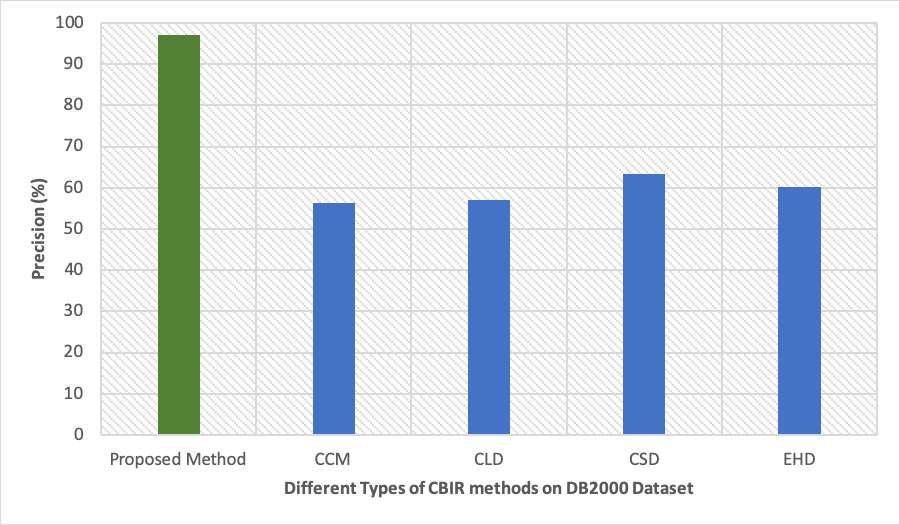}
    \caption{Comparison of average precision between our proposed method and Bose et al.\cite{bose} methods for scope of 20 on DB2000.}
    \label{fig:2000_compare}
\end{figure}

For DB2000, we select Bose et al.'s paper\cite{bose} as the baseline result. This paper\cite{bose} extracted features from the images in two ways: features from colour co-occurrence matrix and features from MPEG-7. As our paper does not take into account Relevance Feedback\cite{bose}, we are only comparing the precision without relevance feedback of this paper\cite{bose} with ours. Figure \ref{fig:2000_compare} shows that our proposed method outperforms all the methods discussed in\cite{bose}.

\begin{figure}[htp]
    \centering
    \includegraphics[scale=0.7]{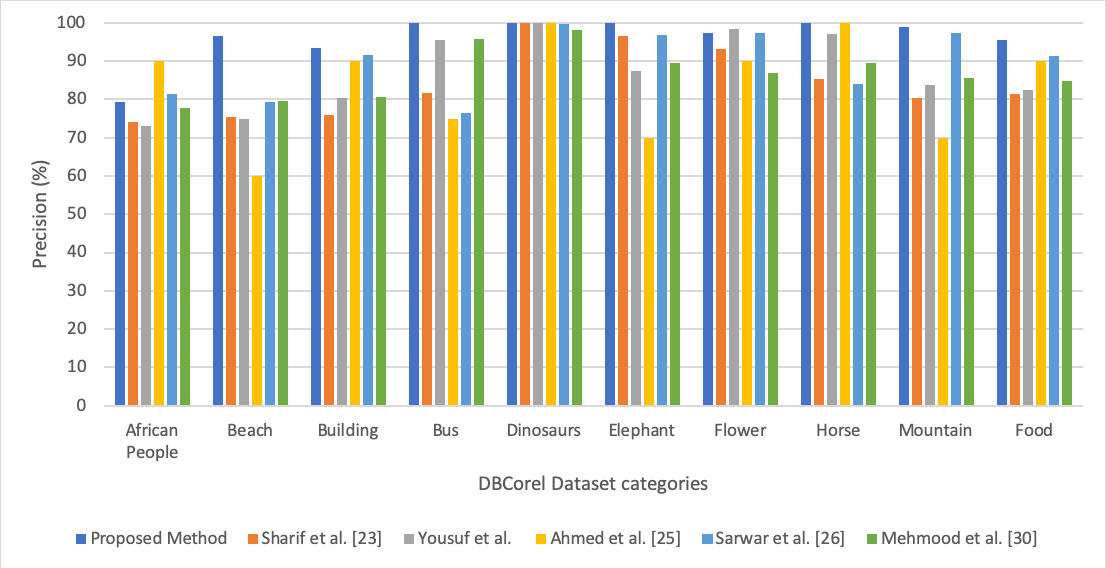}
    \caption{Comparison of category wise precision between our proposed method and recent papers methods for scope of 20 on DBCorel.}
    \label{fig:comp_cat_corel}
\end{figure}

\begin{figure}[htp]
    \centering
    \includegraphics[scale=0.7]{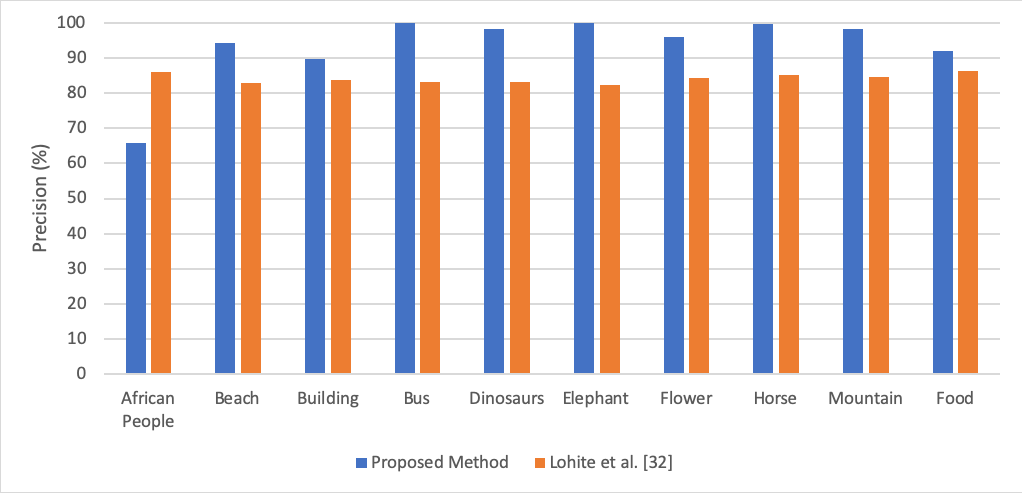}
    \caption{Comparison of category wise precision between our proposed method and Lohite et al.\cite{lohite} methods for scope of 50 on DBCorel. Average precision of our proposed method: 93.39\%. Average precision of Lohite et al.\cite{lohite}: 84.226\%}
    \label{fig:comp_cat_corel50}
\end{figure}

In recent years, many researchers have worked\cite{khokhar, ashraf, sharif, yousuf, ahmed, sarwar, ahmed2, ashraf2, rashno, Mehmood, lohite, ayoob} on DBCorel Dataset extracting different kinds of features and similarity distances. We present two types of comparison with these recent papers on DBCorel Dataset: Category wise precision comparison (Figure \ref{fig:comp_cat_corel}) and average precision comparison (Table \ref{tab:table2}). It shows that our proposed method outperforms all other methods published in the recent papers. Lohite et al.\cite{lohite} calculated category wise precision for scope of 50 instead of 20. We did a comparative study with this algorithm too in Figure \ref{fig:comp_cat_corel50} and showed that except African People category our proposed method works better even for scope 50. 

\begin{table}[ht]
    \centering
    \begin{tabular}{|c|c|c|}
        \hline
        \textbf{Methods} & \textbf{Average Precision (\%)} \\
        \hline
        \textbf{Proposed Method} & \textbf{96.115} \\
        \hline
        Ashraf et al.\cite{ashraf} & 73.5 \\
        \hline
        Sharif et al.\cite{sharif} & 84.39 \\
        \hline
        Yousuf et al.\cite{yousuf} & 87.3 \\
        \hline
        Ahmed et al.\cite{ahmed2} & 83.5 \\
        \hline
        Sarwar et al.\cite{sarwar} & 89.58 \\
        \hline
        Ahmed et al.\cite{ahmed} & 76.5 \\
        \hline
        Ashraf et al.\cite{ashraf2} & 82 \\
        \hline
        Rashno et al.\cite{rashno} & 65.95 \\
        \hline
        Mehmood et al.\cite{Mehmood} & 87.85 \\
        \hline
        Khokhar et al.\cite{khokhar} & 94.3 \\
        \hline
        Ahamed et al.\cite{ayoob} & 82 \\
        \hline
    \end{tabular}
    \caption{Comparison of average precision between our proposed method and recent papers’ methods for scope of 20 on DBCorel.}
    \label{tab:table2}
\end{table}

\begin{figure}[htp]
    \centering
    \includegraphics[scale=0.6]{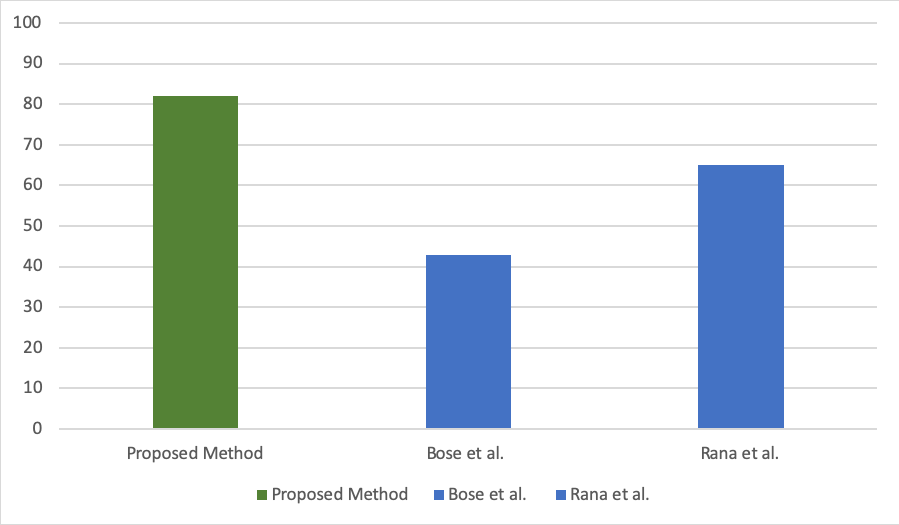}
    \caption{Comparison of average precision among our proposed method, Bose et al.\cite{bose}  and Rana et al.\cite{rana} methods for scope of 20 on DBCaltech.}
    \label{fig:caltech_compare}
\end{figure}

For DBCaltech (Caltech101) Dataset, we produce the comparisn with two algortihms given in \cite{bose, rana}. Figure \ref{fig:caltech_compare} compares the average precision. Multiple CBIR techniques are described in both \cite{bose} \& \cite{rana}. Hence, we picked the average precision of the best methods. 
Clearly the proposed method outperformed both of the other methods.

\section{Real Time CBIR}

We have already seen that due to introduction of very deep neural network model (InceptionResNetV2), our results have been improved quite a significant amount, but it arises the retrieval time of images as a matter of question. Whatever models we use, our ultimate goal is to retrieve images in real-time. In this chapter we will discuss about the time complexity of our CBIR system and will show that in spite of introducing deep models, our system can retrieve images in real time for DBCaltech dataset. Also, we will show that introducing the application of Principal Component Analysis\cite{shlens} will make the image retrieval even faster without sacrificing the precision.

Because of small image count, we did not calculate image retrieval time for DB2000 and DBCorel, as it will be always low. 

\subsection{Principal Component Analysis}

Principal Component Analysis\cite{shlens} (PCA), is a dimensionality-reduction method generally used to reduce the dimensionality of large data-sets, by converting large set of variables into smaller ones which contains most of the information of the original dataset.

Reducing the number of variables of a data set truncates the information of it, but the trick in dimensionality reduction is to trade a little information for simplicity. The reason behind is smaller data sets are easier to handle and visualize. Analyzing data becomes much easier and faster for machine learning algorithms in small sized datasets.

So, to sum up, the idea of PCA is simple—reduce the number of variables of a data set, while preserving as much information as possible.

\subsection{PCA on the Encoded Features}

The encoded feature vector dimension for InceptionResNetV2 is 1536 which seems to be large. So, we did Principal Component Analysis on the 1536 feature vector to reduce its dimension and chose the number of principal components (M) for which the average precision value is maximum. For DBCaltech dataset we are taking roughly 100 PCs to calculate the precision. It is seen that taking the first handful number of PCs results almost the same or sometime better average precision with respect to the whole 1536 features. Average precision with PCA: 82.54\% and average precision without PCA: 82.02\%. In summary PCA increases precision and saves computational time as we are handling with a very reduced dimension.

\subsection{Approach}
Here, we calculate the average query image retrieval time for the scope of 20 on DBCaltech Dataset. This experiment is done with the following combinations: with PCA and without PCA.

To explain the experiment with PCA, at first, we will feed all of our database images through CBIR model (InceptionResNetV2 without the last softmax layer) and PCA respectively, then store those extracted features of dimension 100 of each image in memory as a feature bank. Now when a query image comes it will be passed through CBIR model and PCA respectively. Then we will compare the extracted features from the query image with each of the feature list in the feature bank and ultimately retrieve those images whose features are closer to the query image features evaluated by some similarity metrics i.e. Manhattan Distance, Euclidean Distance etc. So, the time between the feeding of query image and retrieving similar images is the image retrieval time and Fig 41 shows this average image retrieval time. We use the term “average”, because we used all the images for our database as query image and calculated retrieval time for each of these images and finally took mean. This process is test on two machines: 

\textbf{Our local machine}
\begin{itemize}
    \item 1.8GHz Intel Core i5 processor 
    \item 8GB LPDDR3 RAM
\end{itemize}

\textbf{GPU Machine}
\begin{itemize}
    \item GPU: 1 NVIDIA Pascal GPU 
    \item CUDA Cores: 2,048 
    \item Memory Size: 16 GB GDDR5 
    \item H.264 1080p30 streams: 24 
    \item Max vGPU instances: 16 (1 GB Profile) 
    \item vGPU Profiles: 1 GB, 2 GB, 4 GB, 8 GB, 16 GB 
    \item Form Factor: MXM (blade servers) 
    \item Power: 90 W (70 W opt) 
    \item Thermal: Bare Board
\end{itemize}

\begin{figure}[htp]
    \centering
    \includegraphics[scale=0.5]{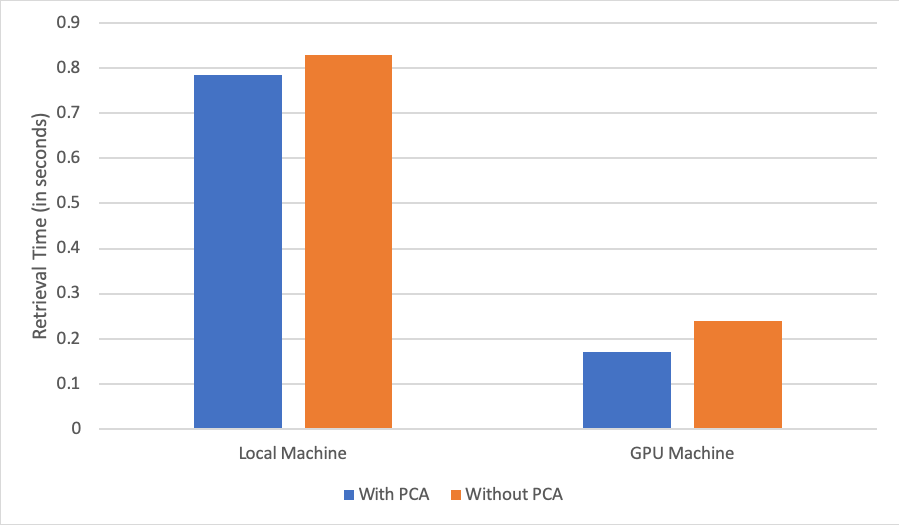}
    \caption{Tested image retrieval time on DBCaltech}
    \label{fig:avg_CBIR_time}
\end{figure}

As we all know that GPUs are highly specialized in parallel computing, so the time required for image retrieval is very less in GPU compared to our local machine. This can be clearly seen in Figure \ref{fig:avg_CBIR_time}. Also it can be seen that using PCA has reduced down the retrieval time a bit.

DBCaltech has 9144 images with 1536 dimensional features (without PCA) and DB2000 has 2000 images with 1536 dimensional features. We can see that our GPU machine and somewhat our local machine also can retrieve images from these dataset in real time. Image retrieval time depends on both the Database size and dimension. Dimension is almost same as long as we use same architecture (InceptionResNetV2 in our case). But if we use a dataset of very high number of images (say millions) then the image retrieval time increases naturally. In case, where we see that searching through all of the database for relevant images is taking much time then we could use a random sample of size say 10,000 or 20,000 to retrieve 20 images from the database of size millions or billions.  

\section{Fast Image Retrieval with Image Clustering}

As mentioned above that as the size of the database increases the image retrieval time also increases. So, we have thought of a novel method to further improve the image retrieval time. This method does clustering of the database images and then only search for images within a specified cluster. 

\subsection{Approach}

The pre-trained model: InceptionResNetV2 we have used during our pre-specified CBIR method was originally trained to predict 1000 classes. Earlier we omitted the last softmax layer and chose the last dense layer for feature extraction. This time we will use both the last dense layer output and softmax layer probability output. The method is explained step by step below. This method has been applied on the Caltech Dataset.

\begin{figure}[htp]
    \centering
    \includegraphics[scale=0.8]{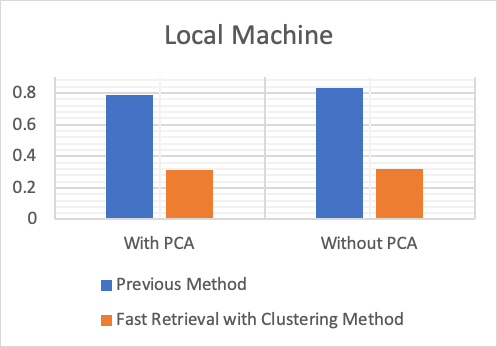}
    \includegraphics[scale=0.8]{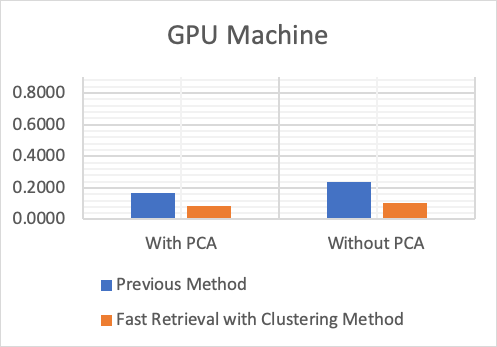}
    \caption{Tested image retrieval time on DBCaltech between proposed Fast Retrieval method and previous method}
    \label{fig:fast_CBIR_time}
\end{figure}

\begin{enumerate}
    \item First, we calculate the last dense layer feature extraction and also calculate the probabilities of assigning to each of the 1000 pre-specified classes for each of the images in the database.
    \item Now according to the probabilities, we assign the each of images to the top 5 classes. For example, let’s say Image\_2.jpg has the probability to be assigned to class 2 with 0.4 probability, class 78 with 0.2 probability, class 9 with 0.15 probability, class 324 with 0.1 probability, class 639 with 0.05 probability and so on with decreasing probabilities, then we assign Image\_2.jpg to class 2, class 78 class 9, class 324 and class 639. We also save the last dense layer feature values for each image in the database in memory.
    \item At the time of retrieval, we calculate both the last dense layer feature values and class probabilities for the query image as well and assigned the query image to top 5 classes based on the probabilities. Let’s say assigned three classes for the query image: query\_1.jpg are class 11, class 258, class 750, class 54 and class 23. 
    \item We accumulate all the images which has any of these classes in their top five classes and calculate similarity measure with the last dense layer feature dimension and retrieve the similar images with only in the accumulated images. 
    \item This gives us a much fast retrieval with respect to the previous method as now we are searching relevant in only a small subset of the whole database (9144 images) instead of searching in the whole 9144 images. For searching in the top 5 classes the average number of images to be searched for any image retrieval becomes only 468.
    \item The reduction in the image retrieval time is shown in Figure \ref{fig:fast_CBIR_time}. This experiment has been tested for with and without PCA on both Local and GPU Machine. From the time we feed the query image to the system to till we get the retrieved images is the image retrieval time of an image. This process is repeated treating each of the images of the database as query image. We noted down the image retrieval time of each of the images and finally took the mean to calculate the mean image retrieval time. 
\end{enumerate}

Note: The precision value of the with this above fast retrieval method becomes 81.48\% while with the previous method it was 82.02\%. So, it is clearly seen that the precision value does not reduce much while the image retrieval time reduces around 2.5 times. The reason behind almost similar results even if searching in the small subset is that the InceptionResNetV2 model predicts the similar images to the same classes, so the clusters of similar images are formed in the classes.

\section{Conclusion}
This paper shows that using pre-trained deep learning features gives better precision result with respect to the features derived by traditional methods e.g. CCM, wavelet etc. 

However, this result can be improved for a specific dataset by introducing the user feedback which is called as Relevance Feedback. Relevance Feedback is basically the feedback from users after each retrieval regarding which results are relevant to the query images and which are not. Using this feedback the CBIR system will start learning and will improve the result gradually.  

There is one limitation of features derived by deep learning is that these features are not rotation-invariant. This means that if we try to retrieve similar images given a same query image but with different orientation angles at every time, the retrieval results will change significantly. Building a rotation-invariant CBIR system may be a next step of improvement over this approach. 

\clearpage
\printbibliography
\end{document}